\documentstyle[aps,pre,amsfonts,preprint]{revtex}
\begin{document}
\draft

\title{Geometry of dynamics and phase transitions
in classical lattice $\varphi^4$ theories}

\author{Lando Caiani$^{1,}$\cite{lando}, Lapo Casetti$^{2,3,}$\cite{lapo},
Cecilia Clementi$^{1,}$\cite{cecilia}, \\
Giulio Pettini$^{4,}$\cite{giulio},
Marco Pettini$^{5,}$\cite{marco} and Raoul Gatto$^{3,}$\cite{raoul}}
\address{$^1$International School for Advanced Studies (ISAS/SISSA), 
via Beirut 2-4, I-34014 Trieste, Italy \\
$^2$Scuola Normale Superiore, Piazza dei Cavalieri 7, I-56126 Pisa, Italy \\
$^3$D\'epartement de Physique Th\'eorique, Universit\'e de Gen\`eve,
24 Quai Ernest-Ansermet, CH-1211 Gen\`eve, Switzerland \\
$^4$Dipartimento di Fisica, Universit\`a di Firenze, Largo E. Fermi 2, 
I-50125 Firenze, Italy \\
$^5$Osservatorio Astrofisico di Arcetri, Largo E. Fermi 5, I-50125 Firenze,
 Italy}

\date {\today}

\maketitle

\begin{abstract}
We perform a microcanonical study of classical lattice
$\varphi^4$ field models in $3$ dimensions with $O(n)$
symmetries. The Hamiltonian flows associated to these systems that undergo a 
second order phase transition in the thermodynamic limit are here 
investigated. The microscopic Hamiltonian dynamics neatly reveals the presence
of a phase transition through the time averages of conventional 
thermodynamical observables. Moreover, peculiar behaviors of the largest 
Lyapounov exponents at the transition point are observed. 
A Riemannian geometrization of
Hamiltonian dynamics is then used to introduce other relevant observables, that
are measured as functions of both energy density and temperature.
On the basis of a simple and abstract geometric model, we suggest that the
apparently singular behaviour of these geometric observables might probe a
major topological change of the manifolds whose geodesics are the natural
motions.
\end{abstract}

\pacs{PACS numbers(s): 05.20.-y; 05.45.+b; 02.40.-k}

\narrowtext

\section{Introduction}

The general problem of the relevance of microscopic dynamics 
to the statistical behaviour of physical systems 
dates back to Boltzmann's ideas at the very beginning of statistical
mechanics, and is still far from being clarified and solved. Within this
framework, one can extract a less general but still challenging question, i.e.
whether the microscopic Hamiltonian dynamics displays some relevant
change when a given system undergoes a phase transition.

Studying  microscopic Hamiltonian dynamics means that -- instead of using 
{\it ensemble} statistical averages -- one numerically computes {\it time} 
averages of the relevant observables. There are two main reasons for so doing:
{\it i)} there exist interesting observables that are intrinsically dynamical,
as is the case of Lyapounov exponents;
{\it ii)} through a differential-geometric description of the dynamics, 
based on simple tools of Riemannian geometry,   
new concepts and methods come to enrich the standard approaches 
to the study of phase transitions, hinting to a possibly deeper 
characterization of their very nature from the standpoint of the 
mathematical structures involved.

The geometric formulation of the dynamics of many-degrees-of-freedom systems 
was first used by Krylov in his pioneering studies on the dynamical
foundations of statistical mechanics \cite{Krylov}. Then, during the last two decades, there have been some attempts to cope with the ergodicity of 
Hamiltonian systems through a 
geometric theory of dynamics \cite{other_works}. A more recent series 
of papers
\cite{Pettini,CasettiPettini,CerrutiPettini,PettiniValdettaro,prl95,CCP},
instead of dealing with ergodicity, successfully addresses the problem of 
explaining and quantifying Hamiltonian chaos within a geometric framework 
where natural motions are seen as geodesics of a suitable Riemannian
manifold (henceforth referred to as ``mechanical manifold'').
Here chaotic dynamics stems from curvature fluctuations
along the geodesics, through a mechanism similar to the parametric 
destabilization of the stable orbits of a pendulum. At variance with a
widespread belief, negative curvatures 
do not appear essential to produce chaos, positive and fluctuating 
curvatures can work as well. 
A very interesting point is that the average degree of instability
of the dynamics is given in terms of curvature-related quantities integrated
over the whole ``mechanical manifold''. This establishes a link between
a {\em dynamical} aspect of a given system --- the stability/instability
of its trajectories --- and some {\em global geometric} 
properties of its associated ``mechanical manifold''.

Now, when a model system displays a phase transition, a natural question arises:
what kind of relationship exists -- if any -- between all the well-known 
major thermodynamic changes occurring at the transition point and the mentioned
global geometric characteristics of the ``mechanical manifolds''?
The present work actually  shows 
that a second order phase transition appears to be associated with an abrupt
change in the global geometry ---  and possibly in the topology, as we
conjecture --- of the ``mechanical manifolds''. 

The above problematics is addressed in the present work by studying the 
dynamics of classical field theories, discretized on a lattice. 
A classical lattice field theory can be
regarded as a discrete classical dynamical system. In particular,
we shall consider the classical $\varphi^4$ theory, whose lattice 
version is a set of coupled nonlinear oscillators. 
 
Equilibrium phase transitions are
usually studied in the framework of the Gibbsian canonical ensemble.
Dynamics, when it is 
considered, is introduced only {\em a posteriori}: the most common procedure
is to describe it by means of non-deterministic equations, usually of
the Langevin type, whose limiting probability distribution is 
the Boltzmann  weight $\exp(-H/T)$ where $H$ is the Hamiltonian
of the system. 
Here we are going to adopt a completely different
approach, i.e. from the very beginning we consider the deterministic
Hamiltonian dynamics, without making explicit assumptions on the equilibrium 
properties of the system, and we observe how the phase transition
is signaled by the dynamics. On rigorous grounds, one cannot be sure that a 
phase transition does exist in a system studied through its
Hamiltonian dynamics, because there is no proof of the fact that the dynamics
is ergodic. Moreover, even assuming that it is ergodic, 
the ergodic measure will be
the microcanonical rather than the canonical one. The two ensembles are
equivalent only in the thermodynamic limit, thus the phenomenology observed
in finite systems, as the systems considered in numerical simulations 
necessarily are, might be different. To give only an example, let us 
consider the phenomenon of {\em ergodicity breaking}, i.e. the fact that
ergodicity is not valid for the whole phase space but only
for disjoint subsets of it. Such a phenomenon is indeed tightly related to 
phase  transitions; in fact, when it occurs it entails a symmetry breaking, as 
in usual phase transitions. But
ergodicity breaking is a more general concept than symmetry breaking,
in fact it is also at the origin of
those phase transitions which do not correspond to the breaking of an
evident symmetry of the Hamiltonian (for example in spin glasses) 
\cite{Goldenfeld,Parisi,ZinnJustin}.
In the canonical ensemble, ergodicity can be broken only in the thermodynamic
limit \cite{LeeYang}, 
while in the microcanonical ensemble, in principle,
there might be ergodicity breaking also in finite systems. Ergodicity 
being a dynamical property, we think that a dynamical
approach is particularly appropriate to study such a phenomenon.

It is worth mentioning here that ergodicity breaking in classical Hamiltonian
systems can be related with supersymmetry breaking \cite{reuter}; this relation
is estabilished within the framework of a path-integral formulation of classical
mechanics, where the bosonic sector of a supersymmetric Lagrangian is given
by a suitable function of the canonical coordinates, obeying standard Hamilton
equations, and the fermionic sector contains ghost fields that - rather 
surprisingly - obey the Jacobi equation describing the stability of classical
paths \cite{gozzi1,gozzi2}. In this framework the spontaneous symmetry breaking 
can occur also at finite volume \cite{reuter}.

Our results show that, as far as
the lattice $\varphi^4$ models considered are concerned, the numerical 
phenomenology, obtained by simulating Hamiltonian dynamics, is perfectly 
consistent with the expectations based on equilibrium statistical mechanics. 
Moreover, we investigate whether the instability of dynamical
trajectories, measured by Lyapounov exponents, is sensitive to the phenomenon
of phase transition \cite{butera}.  In the light of the geometrization of 
dynamics, Lyapounov exponents are also seen as probes of the hidden geometry
of motion, and in fact our results suggest that the deep origin of ergodicity breaking and of the dynamical counterpart of a phase transition could be found
in a major change in the geometric -- or even topologic -- structure of the
``mechanical manifolds'' underlying the dynamics.

The paper is organized as follows:
in Section \ref{sec_phenomenology} we introduce the models studied, we describe 
the numerical techniques which we adopted, and we discuss the phenomenology
of the phase transition as it emerges from the dynamics. In Section 
\ref{geometry} the main definitions and results of the
Riemannian description of Hamiltonian chaos are given, and the behaviour 
of the geometric observables in our models is presented and discussed 
together with an interpretation involving simple topological concepts.
Section \ref{sec_comments} is devoted to some remarks and comments.

\section{Models and numerical results}

\label{sec_phenomenology}

To study the relationship between microscopic dynamics 
and equilibrium phase transitions we consider Hamiltonian systems 
of the standard type 
\begin{equation}
 H [\varphi,\pi ] = \frac{1}{2}\sum_{\bf i} \pi_{\bf i}^2 + 
V(\{\varphi_{\bf i}\}) ~,
\end{equation}
where $\varphi_{\bf i}$ and $\pi_{\bf i}$ are canonically conjugated
coordinates and momenta, ${\bf i}$ labels the sites of a $d$-dimensional
cubic lattice, and $V$ is an interaction potential.

More precisely, we consider models which 
can be derived from the paradigm Hamiltonian
\begin{equation}
H[\varphi]=\int\!d^dx\,\left\{\frac{1}{2}\pi^2({\bf x})+ J\frac{1}{2}
[\nabla\!_d\varphi({\bf x})]^2 
- \frac{1}{2} \varphi^2({\bf x})+\frac{\lambda}{4} \varphi^4({\bf x})\right\}.
\label{hfi4c}
\end{equation}
where  $\pi({\bf x})=\delta L[\varphi, 
\dot{\varphi}]/\delta\dot{\varphi}({\bf x})=\dot{\varphi}({\bf x})$ is the 
canonically conjugated momentum density of  $\varphi({\bf x})$, by discretizing
it on a lattice.   By means of the following substitutions
\begin{eqnarray}
\partial\!_\mu\varphi({\bf x})&\rightarrow&\frac{{\displaystyle \varphi({\bf x}
+a{\bf e}_\mu)-\varphi({\bf x})}}{{\displaystyle a}}\,,\nonumber \\[-0.3cm]
&& \label{discr} \\
{\displaystyle \int\!d^dx} & \rightarrow & {\displaystyle 
a^d\sum\nolimits_{\bf i}}~~~,  
\nonumber
\end{eqnarray}
we obtain
\begin{equation}
H[\varphi ,\pi ]=a^d \sum_{\bf i}\left[\frac{1}{2}\pi_{\bf i}^2 + 
\frac{J}{2a^2}\sum_{\mu=1}^d
(\varphi_{{\bf i}+{\bf e}_\mu}-\varphi_{\bf i})^2-\frac{1}{2} 
m^2\varphi_{\bf i}^2+\frac{\lambda}{4}\varphi_{\bf i}^4\right].
\label{hfi4d}
\end{equation}
where $a$ is the lattice spacing, 
${\bf e}_{\mu}$ is the unit vector in the
$\mu$-th direction of the lattice and $\varphi_{\bf i}=\varphi({\bf x_i})$. 
This system shows (at equilibrium) a continuous phase transition with
nonzero critical temperature corresponding to a spontaneous breaking
of the discrete $O(1)$ --- or ${\Bbb Z}_2$ --- symmetry.

We have also considered the vector versions of this lattice $\varphi^4$ model
described by the Hamiltonian
\begin{equation}
H[\varphi ,\pi ]=a^d \sum_\alpha\sum_{\bf i}\left[\frac{1}{2}(\pi_{\bf i}^
\alpha )^2 + \frac{J}{2a^2}\sum_{\mu=1}^d
(\varphi_{{\bf i}+{\bf e}_\mu}^\alpha - \varphi_{\bf i}^\alpha )^2-\frac{1}{2}
 m^2(\varphi_{\bf i}^\alpha )^2\right] + \frac{\lambda}{4}\sum_{\bf i}
\left[ \sum_\alpha (\varphi_{\bf i}^\alpha )^2\right]^2.
\label{hfi4d_n}
\end{equation}
where the index $\alpha$ runs from $1$ to $n$. We have considered, in addition
to $n=1$, $n = 2$, which is the simplest vector case, and $n=4$, which
is the largest value of $n$ that allowed for a complete numerical study
with our computing resources. 
For $n > 1$ the broken symmetry is a continuous one (the potentials 
are respectively invariant  
under planar rotations, $O(2)$, and under the action of the $O(4)$ group). 
Because of the Mermin-Wagner theorem, the interactions being of short range,
the $O(2)$ and $O(4)$ models can have a second order phase transition only on 
three dimensional lattices.

The Hamiltonian dynamics --- and
thus the related dynamical, thermodynamical and geometrical quantities --- 
is studied by molecular dynamics
simulations performed at several values of the
energy density $\varepsilon = E/N$, which is the relevant physical
parameter as long as our systems are in a microcanonical 
ensemble\footnote{The qualitative features of the results are not
affected if we consider the temperature (average kinetic energy per
degree of freedom) as the physical parameter.}.

\subsection{Numerical study of dynamics and thermodynamics}

The canonical equations of motion
\begin{eqnarray}  
\dot{\varphi}^\alpha_{\bf i}&=&\frac{\displaystyle \partial H}{\displaystyle 
\partial\,\pi^\alpha_{\bf i}} 
\\[0.5cm]
\dot{\pi}^\alpha_{\bf i}&=&-\frac{\displaystyle \partial H}{\displaystyle 
\partial \varphi^\alpha_{\bf i}}
\label{eqcan} \\
\noalign{\hbox{yield}} \nonumber \\[-0.4cm]
&&\dot{\varphi}^\alpha_{\bf i} = \pi^\alpha_{\bf i} \nonumber \\[-0.2cm]
&& \label{eqmoto}\\[-0.3cm]
&&\dot{\pi}^\alpha_{\bf i} = A\sum_{\mu=1}^{d}
(\varphi^\alpha_{{\bf i}+{\bf e}_\mu}+
\varphi^\alpha_{{\bf i}-{\bf e}_\mu}) + 
B \varphi^\alpha_{\bf i} - C \Vert\varphi^\alpha_{\bf i}\Vert^2 
\varphi^\alpha_{\bf i}~~, \nonumber \\[0.1cm]
\noalign{\hbox{with}} \nonumber \\[-0.3cm]
&&A = Ja^{d-2} \nonumber \\
&&B = m^2a^d-2Ja^{d-2}d \label{abbrev} \\
&&C = \lambda a^d~~, \nonumber
\end{eqnarray}
and $\Vert\varphi^\alpha_{\bf i}\Vert^2 =\sum_\alpha(\varphi^\alpha_{\bf i})^2$.
In order to guarantee a faithful numerical representation of a Hamiltonian 
flow, it is necessary that the algorithm updates the canonical
coordinates, $[\varphi^\alpha_{\bf i}(n\Delta t),(\pi^\alpha_{\bf i}(n\Delta t)
]\rightarrow[\varphi^\alpha_{\bf i}((n+1)\Delta t),(\pi^\alpha_{\bf i}((n+1)
\Delta t)]$, by means of
a canonical, i.e. symplectic, transform. Symplectic algorithms ensure the
conservation of Poincar\'e geometric invariants, and, in particular, of
phase space volumes and energy conservation.
We used a very efficient and precise third order symplectic algorithm recently 
proposed \cite{Casetti}, keeping 
the fluctuations of relative energy at $\Delta E/E\simeq 10^{-9}$. 
All the simulations have been performed using words of $64$ bits.
We have always chosen random initial conditions at equipartition among momenta
in order to consider phase space trajectories stemming from initial conditions
that belong to the support of an equilibrium measure.

Along the phase space trajectories --- worked out numerically --- the time
averages of any observable $A$ is computed as
\begin{equation}
\overline{A}^t = \frac{1}{t}\int_0^t\, d\tau A[\pi (\tau ),\varphi (\tau )]~~.
\label{timeaver}
\end{equation}
By means of such averages both dynamical and thermodynamical properties of
the system under investigation can be determined.

One of the most relevant properties of the dynamics is its degree of
instability, because it is related to the efficiency of phase mixing.
Let us remember that the the strength of dynamical instability, i.e. of
{\it chaos}, is measured by the largest Lyapounov exponent $\lambda_1$. 
If we denote by $\cal M$ the phase space of the system and by $X$ a vector 
field on it, such that 
\begin{equation}
\dot x^i = X^i(x^1\dots x^N)
\label{eq1}
\end{equation}
are the equations of motion, a complete integral of this 
 dynamical system defines a one-parameter group of diffeomorphisms of 
$\cal M$, that is $\phi^t: {\cal M}\rightarrow {\cal M}$. Denote by
\begin{equation}
\dot\xi^i = {\cal J}^i_k[ x(t)]\, \xi^k
\label{eq2}
\end{equation}
the tangent dynamics equation, i.e. the realization of the mapping
$d\phi^t: T_x{\cal M}\rightarrow T_{\phi^t(x)}{\cal M}$, where
$[{\cal J}^i_k]$ is the Jacobian matrix of $[X^i]$, then the largest Lyapounov
exponent $\lambda_1$ is defined by
\begin{equation}
\lambda_1 = {\displaystyle\lim_{t\rightarrow\infty}}~\frac{1}{t}\ln \frac
{\Vert\xi (t)\Vert}{\Vert\xi (0)\Vert}
\label{eq3}
\end{equation}
and, by setting $\Lambda [x(t),\xi (t)]= \xi^T\,{\cal J}[x(t)]\, 
\xi /\,\xi^T\xi\equiv\xi^T {\dot\xi} /\xi^T \xi 
=\frac{1}{2}\frac{d}{dt}\ln (\xi^T\xi )$, 
this can be formally expressed as a time average
\begin{equation}
\lambda_1 ={\displaystyle\lim_{t\rightarrow\infty}}~\frac{1}{2t}\int_0^t 
\,d\tau \,\Lambda [x(\tau ), \xi (\tau )]~~.
\label{eq4}
\end{equation}
In practice, as we deal with standard Hamiltonians, 
the tangent dynamics (\ref{eq2}) can be written in the form
\begin{equation}
\frac{d^2\xi^i_q}{dt^2} +\left( \frac{\partial^2 V}{\partial\varphi_i
\partial\varphi^j}\right)_{\varphi (t)} \,\xi^j_q =0
\label{tan_dyn}
\end{equation}
which, integrated along any numerical trajectory of Eqs.(\ref{eqmoto}), 
makes possible the estimate of $\lambda_1$ from
\begin{equation}
\lambda_1(t_{\cal N})=\frac{1}{{\cal N}\Delta t}\sum_{n=1}^{\cal N}\ln
\left(\frac{ \Vert\xi(t_n)\Vert}{ \Vert\xi(t_{n-1})\Vert}\right) ~~,
\end{equation}
where $\{\xi^i\} =(\{\xi^i_q\},\{ \xi^i_p\})$, 
$\xi^i_p(t)=[\xi^i_q(t+\Delta t)-\xi^i_q(t-\Delta t)]/2\Delta t$, and
$t_n=n\Delta t$ ($\Delta t$ is some time interval). The average is extended
 up to a final time $t_{\cal N}$ such that $\lambda_1(t_{\cal N})$ has 
attained a {\it bona fide} asymptotic value.

For what concerns thermodynamic observables, temperature 
--- the basic quantity ---
is determined through the time average of kinetic energy per degree of freedom
\begin{equation}
\frac{1}{2} T = \frac{1}{t}\int_0^t\,d\tau \left\{ \frac{1}{Nn}
\sum_{\alpha ,
{\bf i}}\frac{1}{2}[\pi^\alpha_{\bf i}(\tau )]^2\right\}
\label{temperatura}
\end{equation}
where $N$ is the number of lattice sites
and $t$ is the total time during which a phase space trajectory is
followed. This quantity shows a fast convergence in time and
is expected to differ from its canonical counterpart by
a ${\cal O}(\frac{1}{Nn})$ correction.

Besides the bifurcation of the order parameter --- $\langle\varphi\rangle$ ---
at some critical value of the temperature, a second order phase transition
is signaled by a singular temperature dependence of the specific heat, 
and therefore the 
microcanonical computation of the constant volume specific heat $C_V$ deserves 
special care. An efficient numerical method to compute $C_V$ is devised by
inverting a general formula relating canonical and microcanonical averages
of the squared fluctuations of a generic observable \cite{LPV} by applying it 
to the fluctuations of kinetic energy
\begin{equation}
\overline{\delta K^2}=\widehat{\delta K^2}-\frac{\beta^2}
{C_V}\left(\frac{\partial\widehat{K}}{\partial\beta}\right)^2~,
\end{equation}
where  $C_V=(\partial E/\partial T)$; overbar and hat stand for microcanonical
and canonical averages respectively.

The quantity $\overline{\delta K^2}$ can be easily computed along the 
numerical trajectories, whereas the analytic expressions 
 $\bar{K}=\widehat{K}=N/2\beta$, $\widehat{\delta K^2}=N/(2\beta^2)$
are readily found. By inverting the equation above one immediately finds
a formula for a microcanonical estimate of the canonical specific heat 
\begin{equation}
C_V=\frac{N n}{2}\frac{1}{1-(Nn/2)[(\overline{\delta K^2})/(\bar{K}^2)]}
\label{formulalpv}
\end{equation}
which requires the numerical computation of time averages of kinetic energy
and of its squared fluctuations; $Nn$ is the total number of degrees of freedom.

\subsection{Dynamical evidence of the phase transition}

\subsubsection{Detecting the transition: Binder cumulants}

In the canonical ensemble, a phase transition may show up 
only in the thermodynamic limit. As long as $N$ is finite, 
all the thermodynamic quantities are regular functions of 
the temperature, and ergodicity and symmetry are not broken.
Nevertheless, some marks of the transition show up neatly 
also in a finite system. The specific heat does not diverge, 
but exhibits a peak --- whose height grows with the size of 
the system --- at a temperature $T_c^{C_V}(N)$. In principle 
the order parameter is expected to vanish on the whole 
temperature range for any finite value of $N$, though 
in practice, e.g., in a 
canonical MonteCarlo simulation where the length of the 
sampling of $\varphi$ is necessarily finite, the system is 
trapped in one of the two phases for a ``time'' which grows
exponentially with $N$ \cite{Goldenfeld}, and thus a 
fictitious symmetry breaking is observed at a temperature 
$T_c^{\varphi}(N)$. This temperature, in general, does 
not coincide with $T_c^{C_V}(N)$, even if
\begin{equation}
\lim_{N \to \infty} T_c^{C_V}(N) = \lim_{N \to \infty} 
T_c^{\varphi}(N) = T_c^\infty \,\, .
\end{equation}
In the microcanonical ensemble ergodicity breaking may occur 
also at finite $N$, hence we can expect that a ``true'' 
critical energy exists also at finite $N$. No 
rigorous theoretical result is at our disposal regarding 
this aspect. Nevertheless, on the basis of asymptotic equivalence
of statistical ensembles, the behaviour of microcanonical thermodynamic 
functions is reasonably expected to be similar to the canonical case, 
at least as $N$ is sufficiently large. Indeed this is what is observed, 
as we shall see in the following. In particular, we expect the 
specific heat to exhibit a peak at a critical energy density 
which is a function of $N$.

In the framework of the statistical theory of critical 
phenomena, by means of the finite-size scaling  analysis \cite{Binder,Dunweg}
the critical properties of the infinite system are inferred from 
the values of the thermodynamic observables in finite samples 
of different sizes. In particular it is possible to locate 
the critical point by means of the so-called {\em Binder 
cumulants} \cite{Binder}. The Binder cumulant $g$ that we have
computed for our systems is defined as
\begin{equation}
g = 1 - \frac{\langle \varphi^4 \rangle}{3\langle \varphi^2 
\rangle^2}\,\, ,
\label{binder}
\end{equation}
where 
\begin{eqnarray}
\langle\varphi^{2n}\rangle & = &\left\langle \left( 
\sum_\alpha\langle\varphi\rangle_\alpha^2\right)^n\right\rangle \nonumber \\
\langle\varphi\rangle_\alpha & = & \sum_{\bf i}\varphi^\alpha_{\bf i}~~.
\nonumber
\end{eqnarray}
In the disordered phase the probability distribution of the 
order parameter will be nearly Gaussian with zero mean, 
hence $g \simeq 0$. At variance to this, at zero temperature (or 
energy), when $\varphi_i \equiv \varphi_0$ with no 
fluctuations, $g = 2/3$. At different sizes of the system, 
$g$ will decay following different patterns $g(N,T)$ 
from $2/3$ to $0$ at increasing temperature.
The remarkable fact is that the 
value of $g$ at $T_c^{\infty}$ is {\em independent} of $N$, 
provided $N$ is large enough for the scaling regime to set 
in; hence the critical point can be located by simply 
looking at the intersection of the different curves $g(N,T)$ 
for different values of $N$. In principle, two different 
sizes are sufficient to locate the transitions; in practice, 
owing to the unavoidable numerical errors which affect $g$, 
it is necessary to consider at least three values of 
$N$. Moreover, the value of $g$ at the critical point, 
usually referred to as $g^*$, is a universal quantity, like the 
critical exponents; for a simple proof see e.g. 
Ref. \cite{Dunweg}. The importance of the Binder cumulant 
method is not only that it allows to easily locate the 
critical temperature, without the need of an extrapolation 
of the asymptotic behaviour of the fictitious finite-$N$ 
critical temperatures, but also that such an estimate of 
$T_c^\infty$ is independent of the other thermodynamic 
observables like $\langle \varphi \rangle$ or $C_V$, and 
this is obviously a great advantage in determining the 
actual critical behaviour --- in particular the critical 
exponents. Moreover, one can regard the existence of a 
crossing of different curves $g(N,T)$ as a ``proof'' for
the existence of a phase transition in the system under 
investigation. This may be useful in various cases where the 
presence of singularities in the thermodynamic functions or 
the existence of a nonzero order parameter are difficult to 
observe (e.g. this is the case of spin glasses \cite{Young}).

The theory behind the Binder cumulant method is totally 
internal to {\it canonical} statistical mechanics: to our 
knowledge no extension of this theory to the {\it microcanonical} 
ensemble exists. Nevertheless we will adopt the pragmatic 
point of view of assuming its validity as a numerical tool 
also in our dynamical simulations, and our operative 
definition of the critical energy density 
$\varepsilon_c^\infty$ will be the intersection point of the 
curves $g(N,\varepsilon)$ at different $N$. The consistency of 
the method will be checked {\em a posteriori}. In the 
following, unless explicitly stated otherwise, 
$\varepsilon_c$ and $T_c$ will denote respectively 
$\varepsilon_c^\infty$ and $T_c^\infty$.

The results for $g(N,\varepsilon )$ at different sizes for the 
$\varphi^4$ lattice models are shown in Figs.\ref{fig1}$a,b,c$. 
The crossing of the various 
curves at $\varepsilon_c \simeq 31$ for the $O(1)$ model is
quite evident, and similarly at 
$\varepsilon_c\simeq 44$ for the $O(2)$ model, and at $\varepsilon_c\simeq 56$
for the $O(4)$ model.

Such estimates of the critical energy densities are obviously far from being
extremely accurate. However, we are mainly interested in 
showing that the dynamical phenomenology is actually consistent with 
the existence of an equilibrium phase transition at finite 
energy density, and the values of $\varepsilon_c$ are needed to 
understand whether or not the singular (or, more generally, 
peculiar) behaviours of the observables 
--- either thermodynamical, or strictly dynamical, or 
geometric ones --- that we are going to study can be 
associated with the phase transition.

\subsubsection{Temperature}

The temperature of the $\varphi^4$ systems, 
numerically determined according to Eq. (\ref{temperatura}), is 
plotted in Fig.\ref{fig2} as a function of the energy density $\varepsilon$.
Note that for all the models a change of the function $T(\varepsilon)$ is
clearly evident at $\varepsilon = \varepsilon_c$.

By plotting the Binder cumulants {\em vs.} the
temperature $T$, the critical values 
$T_c$ are obtained for all the models and are found in complete agreement
with the outcomes of the $T(\varepsilon )$ curves. These values are:
$T_c\simeq 35$ for the $O(1)$ model, $T_c\simeq 25$ for the $O(2)$ model,
and $T_c\simeq 16$ for the $O(4)$ model.

\subsubsection{Specific heat}

The specific heat $c_{_V}=C_V/Nn$ per degree of freedom of the $\varphi^4$ 
models here considered, computed according to Eq. (\ref{formulalpv}), 
is plotted {\it vs.} the temperature in Fig.\ref{fig3}.
The asymptotic values of the specific heat in the limits 
$T \to 0$ and
$T \to\infty$ are exactly known. In fact at low 
energies the anharmonic terms in the Hamiltonian can be 
neglected, thus the system behaves as a collection af harmonic 
oscillators and $c_{_V} \to 1$ as $T\to 0$. In the high-energy limit 
the quadratic terms in the potential are negligible with 
respect to the quartic ones, whence  
$c_{_V} \to 1/2+1/4=3/4$ as $T\to \infty$.\label{limcalspec} At intermediate 
energy densities,  neat peaks show up whose positions are close 
to $T_c$ for each model respectively. The heights of the 
peaks are found to grow with $N$ and to decrease with $n$.

\subsubsection{Dynamical properties}

We have shown that the outcomes of the dynamical numerical
simulations of the scalar and vector versions of the lattice $\varphi^4$
model are perfectly consistent with the expectations of the effects of a second
order phase transition on a finite sample. 
As already motivated above, this first result is non-trivial. 
Up to now its content is that --- for all practical purposes --- a dynamical
simulation is actually equivalent to a microcanonical one, so that, at 
sufficiently large $N$, the results are in natural agreement with canonical 
statistical mechanics.
All these results concern time averages: the time variable,
even if not eliminated from the very beginning as in the statistical
approach, has been nonetheless integrated out in the averaging procedure. 
But we can also wonder what are the properties --- if any --- that are peculiar
to the dynamics and that can be considered relevant to the description of
the phase transition itself.
Moreover we have already noticed that the phenomenon
of ergodicity breaking has a deep dynamical origin, therefore we can try to 
understand what features are associated to a Hamiltonian ergodicity
breaking.
 
The lattice $\varphi^4$ models under investigation are nonintegrable dynamical 
systems. In the two limits $\varepsilon \to 0$ and 
$\varepsilon \to \infty$, these systems become integrable. The 
two integrable limits are respectively those of a system of coupled 
harmonic oscillators and of a system of independent quartic 
oscillators. The dynamics is always chaotic over the 
whole energy range. Nevertheless, in analogy to other 
nonlinear oscillator systems, by varying the energy  we expect that 
qualitatively different dynamical regimes will be found,
characterized by a transition between different behaviours of the largest 
Lyapounov exponent $\lambda_1$ as a function of energy density or, equivalently, 
temperature. This phenomenon is attributed to a dynamical transition between 
weak and strong chaos, it is known 
as Strong Stochasticity Threshold (SST) and is discussed in refs.
\cite{PettiniLandolfi,PettiniCerruti}. 
In particular, the following questions 
naturally arise. Is there any peculiar behaviour of the 
Lyapounov exponent in correspondence with the phase 
transition? Is there a transition between strongly and 
weakly chaotic regimes also in these models, 
and, in the affirmative case, is there any relationship between these 
different dynamical regimes and the thermodynamic phases?

We must say from the very beginning that there are not yet 
conclusive answers to these questions. The study of 
a possible relation between chaos and phase transitions is a 
very recent issue \cite{Ruffo}, and the results so far obtained and reported
in the literature range from the claim of the 
discovery of a ``universal'' divergence in $\lambda_1$ near 
criticality in a class of models 
describing clusters of particles \cite{Rapisarda}, to the 
observation that the Lyapounov exponent attains its minimum 
in correspondence with the phase transition in Ising-like 
coupled map lattices \cite{Duke_pre}, and to the apparent 
insensitivity to the liquid-solid phase transition of the  
Lyapounov spectra of hard-sphere and Lennard-Jones systems 
\cite{Dellago}. 

Our simulation results are plotted in Figs.\ref{fig4} and \ref{fig5}.
The $O(1)$ case has been studied more extensively than the others because
of practical reasons of computational effort (for example single runs for
the $O(4)$ model usually required at least two weeks of CPU time on a fast
HP 9000/735 computer).

The first numerical evidence is that in presence of a second order
phase transition a rather sharp and ``cuspy'' transition between different
behaviours of $\lambda_1(T)$ is found at $T_c$ (where the critical values $T_c$
are those determined by means of Binder cumulants).
Moreover, the qualitative behaviour of $\lambda_1(T)$ appears very
different in the thermodynamically ordered and disordered
regions respectively. In fact, in the former region $\lambda_1$ rapidly 
increases with $T$, whereas in the latter region $\lambda_1(T)$ displays
an almost flat pattern above $T_c$ (note that $\lambda_1(T)$ is expected to
change again at very large $T$ because the dynamics is asymptotically
integrable in the limit $T\to \infty$; this effect has been numerically checked
at very high temperatures and is clearly evident in  Fig.\ref{fig4} --- $O(1)$ 
case --- at $T/T_c\sim 10^4$).
This suggests that the phase transition has a dynamical counterpart in a 
passage from a weakly to a strongly chaotic regime. 

It is remarkable that the shape of $\lambda_1(T)$ is significantly different
in presence or in absence of a second order phase transition. In fact, in
the case of one dimensional lattices with short range interactions --- where 
no phase transition is present --- $\lambda_1(T)$ has a very smooth pattern
(see Ref.\cite{CCP}). This fact has been checked more
specifically for the $\varphi^4$ model by computing $\lambda_1(T)$ for the
$O(2)$ symmetry case on a two-dimensional lattice; as a consequence of the
Mermin-Wagner theorem, here a second order phase transition is forbidden and
in fact this model undergoes an infinite order phase transition 
(Kosterlitz-Thouless-Berezinsky). 
The shape of $\lambda_1(T)$ again displays a major
change so that the low and high temperature regimes are very different.
However the transition between these two regimes is now smooth \cite{cccp}.

It is worth to emphasize that the average of a {\em local} property of
microscopic dynamics --- the average instability measured by $\lambda_1$ --- 
is sensitive to a {\em collective} phenomenon like a second order phase 
transition.

It could be argued that in the critical region almost any ``honest'' observable
will show a peculiar behaviour and that this reflects the tendency of the
statistical measure to become singular at the transition point, regardless
of the ensemble chosen.
In the framework of equilibrium statistical mechanics this is certainly true, 
because Gibbs measure is the fundamental mathematical object upon which
everything relies. In the thermodynamic limit also the microcanonical measure,
which is the invariant measure of the microscopic Hamiltonian dynamics, will
have to become singular. However the microcanonical measure is not the 
ultimate mathematical entity that can be considered, so that the Hamiltonian
dynamics approach gives meaning to the question of the possible existence of a 
{\it more fundamental} phenomenon at the very ground of a phase transition.

Lyapounov exponents provide the necessary link to such unexplored land.
The details on this point are given in the next Section, where we recall
how the geometrization of Hamiltonian dynamics proceeds in the language of
Riemannian geometry, and how average geometric properties of some suitable 
manifold directly influence the average dynamical instability quantified by
$\lambda_1$.
\smallskip
\section{Geometry of dynamics and the phase transition}
\label{geometry}
\smallskip
Let us here sketch the main points of the Riemannian theory of chaos in 
physical systems, details can be found in ref.s 
\cite{Pettini,CasettiPettini,CerrutiPettini,PettiniValdettaro,prl95,CCP}. 

\subsection{Riemannian geometrization of newtonian dynamics}
\label{sec_geometry}

The trajectories of a dynamical system described by the Lagrangian function
\begin{equation}
L({\bf q},{\bf\dot q})={1\over 2}a_{ik}({\bf q})\dot q^i\dot q^k -V({\bf q})
\label{lagr}
\end{equation}
are geodesics of the configuration
space endowed with a proper Riemannian manifold structure described by the 
metric tensor 
\begin{equation}
g_{ik}({\bf q})=2[E-V({\bf q})]a_{ik}({\bf q})~.
\label{Jmetric}
\end{equation}
This metric is known as Jacobi metric and is defined in the region of the
configuration space where $E>V({\bf q})$.
In local coordinates, the geodesic equations on a Riemannian manifold are
given by
\begin{equation}
{{d^2 q^i}\over{ds^2}}+ \Gamma^i_{jk}{{dq^j}\over{ds}}{{dq^k}\over{ds}}=0
\label{geo} 
\end{equation}
where $s$ is the proper time and $\Gamma^i_{jk}$ are the Christoffel 
coefficients of the Levi-Civita connection associated with $g_{ik}$, i.e.
$\Gamma^i_{jk}={1\over{2 W}}\delta^{im}(\partial_jW\delta_{km}+\partial_kW
\delta_{mj}-\partial_mW\delta_{jk})$, where  $W=E-V({\bf q})$; 
proper time and physical time are related by $ds^2=2W^2dt^2$.
By direct computation, using $g_{ik}=(E-V({\bf q}))\delta_{ik}$,
it can be easily verified that the geodesic equations
yield
\begin{equation}
{{d^2q^i}\over{dt^2}}= -{{\partial V}\over{\partial q^i}}\label{newton}
\end{equation}
i.e. Newton's equations associated with the Lagrangian (\ref{lagr}).
These equations can be also derived as geodesics of a manifold consisting of 
an enlarged configuration spacetime 
$M\times {\Bbb R}^2$, with local coordinates
$(q^0,q^1,\ldots,q^i,\ldots,q^N,q^{N+1})$. To such purpose this space is 
endowed with a non-degenerate pseudo-Riemannian metric, first 
introduced by
Eisenhart \cite{Eisenhart}, whose arc-length is 
\begin{equation}
ds^2 = g_{\mu\nu}\, dq^{\mu}dq^{\nu} = 
a_{ij} \, dq^i dq^j -2V({\bf q})(dq^0)^2 
+ 2\, dq^0 dq^{N+1} ~,
\label{g_E}
\end{equation}
called {\em Eisenhart metric}. 
The natural motions are obtained as the canonical projection 
of the geodesics
of $(M\times {\Bbb R}^2,g_E)$ on the configuration 
space-time, 
$\pi : M\times {\Bbb R}^2 \mapsto M\times {\Bbb R}$. 
Within the totality of geodesics only those whose 
arclength is 
positive-definite and is given by $ds^2 = c_1^2 dt^2$
correspond to natural motions, what is equivalent to requiring the 
condition $q^{N+1} = \frac{1}{2}c_1^2 t + c^2_2 - \int_0^t L\, 
d\tau~$ for the extra-coordinate $q^{N+1}$ \cite{Pettini,CasettiPettini};
$c_1$ and $c_2$ are real arbitrary constants.

\subsection{Curvature and instability of geodesic motions}

There is an important relation between the curvature of a 
manifold and the stability of its geodesics. It is described by the
Jacobi -- Levi-Civita (JLC) equation for the {\em geodesic
separation vector field} $J(s)$. 

The evolution of $J$ contains the whole information on the 
stability
--- or instability --- of a given reference geodesic 
$\gamma (s)$: 
in fact if $|J|$ grows exponentially then the geodesic will 
be unstable
in the Lyapounov sense, otherwise it will be stable. It is 
remarkable that
such an evolution is completely determined by the Riemann curvature 
tensor $R^i_{jkl}$ according to the JLC equation
\begin{equation}
\frac{\nabla^2 J^i}{ds^2}\, +\, R^i_{jkl}\frac{dq^j}{ds}J^k\frac{dq^l}{ds}
=\,0~,\label{eq_jacobi}
\end{equation}
where $\frac{\nabla}{ds}$ is the covariant derivative.

In the large $N$ case, under suitable hypotheses \cite{prl95,CCP}, it is possible to derive a scalar effective stability equation.
Briefly, among the others, the main assumptions are: 
$i)$ that the ambient manifold is {\it almost-isotropic}, this  
essentially means that -- after some suitable coarse-graining -- 
the ambient manifold would look like a constant curvature manifold;
$ii)$ that the curvature felt along an unstable geodesic can be reasonably 
modeled by a gaussian stochastic process. 
The final result is \cite{CCP}
\begin{equation}
{{d^2\psi}\over{ds^2}}+ \langle k_R\rangle_\mu\,\psi + 
\langle\delta^2 k_R\rangle_\mu^{1/2}\,\eta (s)\,\psi = 0~,
\label{eq_Hill_psi}
\end{equation}
where $\psi$ denotes any of the components of $J$ in Eq. (\ref{eq_jacobi})
because now all of them 
obey the same effective equation of motion; $\langle k_R\rangle_\mu =
\frac{1}{N}\langle K_R\rangle_\mu$ where $K_R$ is the Ricci curvature of
the ambient manifold: $K_R=R_{ik}\dot q^i \dot q^k$ and $R_{ik}=R^j_{ijk}\,$;
$~\langle\cdot\rangle_\mu$ stands for microcanonical average, and  
$\langle \delta^2 k_R \rangle_\mu$ is a shorthand for  
$\frac{1}{N-1}\langle \delta^2 K_R \rangle_\mu$, the mean square
fluctuation of the Ricci curvature; $\eta (s)$ is a gaussian 
$\delta$-correlated random process of zero mean and unit variance. 

Equation (\ref{eq_Hill_psi}) is a scalar equation which, 
{\em independently of the
knowledge of dynamics}, provides a measure of the average 
degree of instability
of the dynamics through the growth-rate of $\psi (s)$. 
The peculiar
properties of a given Hamiltonian system enter Eq. 
(\ref{eq_Hill_psi}) through
the global geometric properties $\langle k_R\rangle_\mu$ and 
$\langle\delta^2 k_R\rangle_\mu$ of the ambient Riemannian 
manifold.
Moreover $\langle k_R\rangle_\mu$ and 
$\langle\delta^2 k_R\rangle_\mu$ are functions of the energy 
$E$ of the system
--- and of the energy density $\varepsilon = E/N$ as well, which is
the relevant quantity at $N \to \infty$ ---
so that from (\ref{eq_Hill_psi}) we can obtain the energy 
dependence of the 
geometric instability exponent.

Equation (\ref{eq_Hill_psi}) is of the form
\begin{equation}
{{d^2\psi}\over{ds^2}}+ \Omega (s)\, \psi =0
\label{eq_stoc_osc}
\end{equation}
representing a stochastic oscillator where the squared frequency $\Omega (s)$ 
is a stochastic process; the derivation of this  
equation does not depend on a particular choice of the 
metric. For Hamiltonian systems with a diagonal kinetic energy matrix, 
i.e. $a_{ij} = \delta_{ij}$, by 
choosing as ambient manifold for the geometrization of  
dynamics the enlarged configuration space-time equipped with Eisenhart 
metric (\ref{g_E}),
it is found that the only non-vanishing component of the Ricci tensor  
is $R_{00} = \triangle V$, 
thus Ricci curvature is a function of the coordinates $q^i$ only and one has $k_R(q) = \triangle V/N$.
Using $dt^2 = ds^2$, the stochastic oscillator equation (\ref{eq_stoc_osc}) 
can be written 
\begin{equation}
\frac{d^2 \psi}{dt^2} + \Omega (t)\, \psi = 0~,
\label{stoc_osc_t}
\end{equation}
where mean and variance of $\Omega (t)$ are given by 
\begin{eqnarray}
\Omega_0 & = & \langle k_R \rangle_\mu ~ = ~ 
\frac{1}{N}\langle \triangle V \rangle_\mu ~, \label{k_0} \\
\sigma^2_\Omega & = & \langle \delta^2 k_R \rangle_\mu ~ = ~
\frac{1}{N}\left( \langle (\triangle V)^2 \rangle_\mu - 
\langle \triangle V \rangle^2_\mu \right) ~. \label{sigma_k}  
\end{eqnarray}
The process $\Omega (t)$ is specified by $\Omega_0$, $\sigma^2_\Omega$ and
its time correlation function 
$\Gamma_\Omega (t_1,t_2)$. We consider a stationary and $\delta$-correlated 
process $\Omega (t)$ with $\Gamma_\Omega (t_1,t_2) = 
\tau \, \sigma_\Omega^2 \, \delta(\vert t_2 - t_1\vert )~$, where $\tau$ is a characteristic time scale of the process.
At present the evaluation of this time scale is still a rather delicate point, 
where some arbitrariness enters the theory.
In Ref. \cite{CCP} these two time scales are defined by
\begin{equation}
\tau_1  =\left\langle\frac{dt}{ds}\right\rangle 
          \frac{\pi}{2\sqrt{\Omega_0 +\sigma_\Omega}}~~,~~~~
\tau_2 = \left\langle\frac{dt}{ds}\right\rangle 
          \frac{\Omega_0^{1/2}}
         {\sigma_\Omega }~,
\label{tau1}
\end{equation}
that are combined to give $\tau$ as follows
\begin{equation}
\tau^{-1} =  2 \left(\tau_1^{-1} + \tau_2^{-1}\right)~.
\label{taufinale}
\end{equation}
As we shall see below, at low temperatures this formula seems to 
predict a satisfactory temperature-dependence of $\tau$, in fact, by adjusting 
a constant factor that multiplies $\tau_1$, the theoretical prediction 
of $\lambda_1(T)$ is in very good agreement with numerical computations. 
At high temperatures we have to take care of the fact that $\Omega_0$ and
$\sigma_\Omega$ are both increasing functions of $T$, even though the system 
approaches an integrable limit.

Whenever $\Omega (t)$ in Eq. (\ref{stoc_osc_t}) has a non-vanishing stochastic
component, the solution $\psi (t)$ is exponentially growing on the average
\cite{VanKampen}. 
Our estimate for the (largest) Lyapounov exponent is then given by the
growth-rate of $\Vert (\psi ,\dot\psi)(t)\Vert^2 $ according to the definition 
\begin{equation}
\lambda_1 = \lim_{t\to\infty} \frac{1}{2t} \log 
\frac{\psi^2(t) + \dot\psi^2(t)}{\psi^2(0) + 
\dot\psi^2(0)}~.
\label{def_lambda_gauss}
\end{equation}

The ratio $(\psi^2(t) + \dot\psi^2(t))/(\psi^2(0) + \dot\psi^2(0))$ is
computed by means of a technique developed by Van Kampen, summarized in
Ref. \cite{CCP}, that yields 
\begin{eqnarray}
\lambda_1(\Omega_0,\sigma_\Omega,\tau)  & = & \frac{1}{2}
\left(\Lambda-\frac{4\Omega_0}
{3 \Lambda}\right)~, \nonumber \\
\Lambda  & = & 
\left(2\sigma^2_\Omega\tau+\sqrt{\left(\frac{4\Omega_0}{3}
\right)^3+(2\sigma^2_\Omega\tau)^2}\,\right)^{1/3}~.
\label{lambda_gauss}
\end{eqnarray}
The quantities $\Omega_0$, $\sigma_\Omega$ and $\tau$ can be computed
as static, i.e., {\it microcanonical} averages. Therefore Eq. 
(\ref{lambda_gauss}) gives an analytic, though approximate, 
formula for the largest Lyapounov exponent independently of the 
numerical integration of the dynamics and 
of the tangent dynamics. 

\subsection{Geometric signatures of the phase transition}

\label{sec_geom_transition}

As already noted above, on the one hand  the largest
Lyapounov exponent is sensitive to the phase transition, on the other hand, 
in the Riemannian description of chaos, $\lambda_1$ is intimately related to
the average curvature properties of the ``mechanical manifolds''. These 
quantities are computed as integrals on manifolds just like other
statistical quantities of thermodynamic kind. This means that 
by means of statistical-mechanical-like computations we can obtain non-trivial
informations about dynamics.
Hence the following questions: is there any peculiarity in
the geometric properties associated with the dynamics of systems
which --- as statistical systems in thermal equilibrium --- 
exhibit a phase transition?
And in particular, do the curvature fluctuations show any noticeable
behaviour in correspondence with the phase transition itself?

\subsubsection{Results of the computations}

Let us now report on the results of the 
the computation of the geometric properties of
the ``mechanical'' manifolds sampled by the numerical geodesics.

For the $\varphi^4$ models, the Ricci curvature per degree of freedom 
along a geodesic of $(M\times{\Bbb R}^2, g_E)$ is given by
\begin{equation}
k_R = \frac{1}{Nn}\sum_{\alpha =1}^n\sum_{\bf i}\frac{\partial^2V}{\partial 
(\varphi^\alpha_{\bf i})^2} = 2Jd -m^2 + \lambda\frac{n+2}{Nn}\sum_{\alpha =1}^n
\sum_{\bf i}(\varphi^\alpha_{\bf i})^2~.
\label{riccifi4}
\end{equation}
High and low temperature behaviours of this quantity can be easily derived.
In the limit $T\rightarrow 0$ we can replace, at any site ${\bf i}$ of the 
lattice, $\Vert\varphi^\alpha_{\bf i}\Vert^2=\sum_{\alpha =1}^n
(\varphi^\alpha_{\bf i})^2$ with the constant value $\varphi_0^2 =
m^2/\lambda$; this value is obtained by minimizing the potential part
of the Hamiltonian (\ref{hfi4d_n}). Hence, for a generic $O(n)$ case, we have
\begin{equation}
{\displaystyle\lim_{T\rightarrow 0}}\, k_R = \frac{2}{n}(J d n + m^2)
\label{ricciT=0}
\end{equation}
and with the values we chose for the constants --- $J=1$, $m^2=2$, $d=3$, 
$\lambda=0.1$  --- it is
$k_R=10$ in the $O(1)$ case, $k_R=8$ in the $O(2)$ case, and $k_R=7$ in the 
$O(4)$ case respectively. These values perfectly check with our numerical 
findings as is shown by Fig.\ref{fig6} where 
$\kappa (T) =(\langle k_R\rangle_t(T) -2Jd)/(\langle k_R\rangle_t(T=0)-2Jd)$
is synoptically displayed for all the models; the average 
$\langle\cdot\rangle_t$ is defined in Eq. (\ref{timeaver}).
At low temperature $\langle k_R\rangle_t(T)$ only slightly deviates from its
limiting zero-temperature value, as is shown by Fig.\ref{fig7}; this fact 
is intuitively interpreted as a sign of a weakly chaotic dynamics.

Also in the opposite limit, $T\rightarrow\infty$, these systems are again
integrable. In fact at increasing temperature the variables 
$\varphi^\alpha_{\bf i}$ become larger and larger so that the Hamiltonian
(\ref{hfi4d_n}) describes a collection of quartic oscillators that are less and
less perturbed by the quadratic coupling term. In this limit the canonical
partition function is factored in terms of functions of the following form
\begin{equation}
\int_0^\infty dx\, x^\nu e^{-ax^\alpha} = \frac{1}{\alpha}\Gamma \left(
\frac{\nu +1}{\alpha}\right)\, a^{-(\nu +1)/\alpha}~,
\end{equation}
with $\nu =0$, and where 
$\Gamma (x)$ is the Euler gamma. Hence the canonical average 
of any even power $\nu$ of the field is
\begin{equation}
\langle (\varphi^\alpha_{\bf i})^\nu\rangle = \left( \frac{\beta\lambda}{4}
\right) ^{-\frac{\nu}{4}}\, \frac{\Gamma\left(\frac{\nu+1}{4}\right) }
{\Gamma\left(\frac{1}{4}\right) }
\label{powers}
\end{equation}
and vanishes for any odd power $\nu$.

>From Eqs.(\ref{riccifi4}) and (\ref{powers}) we find a canonical estimate of 
$\langle k_R\rangle_\mu$ which differs from the microcanonical one by 
${\cal O}(\frac{1}{N})$ terms 
\begin{equation}
\langle k_R\rangle_\mu \sim 2Jd - m^2 +{\displaystyle
 \frac{2(n+2)\Gamma \left(\frac{3}{4}
\right) \sqrt{\lambda}}{\Gamma \left(\frac{1}{4}\right)}} \sqrt{T}\;+{\cal O}(
\frac{1}{N})~~.
\label{ricciTinf}
\end{equation}
This prediction is compared to the numerically computed values of 
$\langle k_R\rangle_t(T)$ in Fig.\ref{fig7}; at very high temperature the
agreement is very good.

In order to compute the average curvature fluctuations, 
$\langle\delta^2 k_R\rangle_\mu$, we first notice that
\begin{equation}
\langle\delta^2 k_R\rangle =\langle k_R^2\rangle -\langle k_R\rangle^2 =
\lambda^2\left(\frac{n+2}{Nn}\right)^2\left\{
\left\langle \left(\sum_{\bf i}\Vert\varphi_{\bf i}\Vert^2\right)^2\right\rangle
-\left(\sum_{\bf i}\langle\Vert\varphi_{\bf i}\Vert^2\rangle\right)^2\right\}
\label{fluttuazione}
\end{equation}
and, as in the large $T$ limit we consider all the $\varphi^\alpha_{\bf i}$
decoupled, we find
\begin{equation}
\langle\delta^2 k_R\rangle =\lambda^2\,(n+2)^2\left\{ \langle (\varphi^\alpha_
{({\bf i})})^4\rangle - \langle (\varphi^\alpha_{({\bf i})})^2\rangle^2\right\}
\label{fluttuazione1}
\end{equation}
where $\varphi^\alpha_{({\bf i})}$ denotes any representative of the now
independent degrees of freedom. The Gibbsian, canonical, average in the
$T\rightarrow\infty$ limit is now easily found to be
\begin{equation}
\langle\delta^2 k_R\rangle^G\sim\left\{\frac{\Gamma (5/4)}{\Gamma (1/4)} -
\left[ \frac{\Gamma (3/4)}{\Gamma (1/4)}\right]^2\right\}(n+2)^2 4\lambda\, T~.
\label{flut-can}
\end{equation}
In order to compare the predictions of Eq. (\ref{flut-can}) with our numerical
results, and also in order to use it in the analytic prediction of the Lyapounov 
exponent, we have to take into account the correction that relates canonical 
and microcanonical averages \cite{VanKampen} that now reads
\begin{equation}
\langle\delta^2 k_R\rangle_\mu = \langle\delta^2 k_R\rangle^G - \frac{\beta^2}
{C_V}\left( \frac{\partial\langle k_R\rangle}{\partial\beta}\right)^2~~.
\label{flutt-mic}
\end{equation}
The high temperature partition function $Z$ is obtained by raising to the
$Nn$-th power the integral $\int\,d\varphi \exp [-\beta (\lambda/4)\varphi^4]
\sim\beta^{-1/4}$. Then, using $F=-(1/Nn\beta)\ln Z$ and $C_V=-T(\partial^2 F/
\partial T^2)$, we find $c_{_V}\rightarrow 1/4$. 
By the way, this is in very good agreement with our numerical results 
for the high temperature values of $c_{_V}$;
this is somehow less clear in the $O(4)$ case because $c_{_V}$ was computed only
in the transition region. From Eqs. (\ref{flutt-mic}) and (\ref{flut-can}) we
can now obtain the final result
\begin{equation}
\langle\delta^2 k_R\rangle_\mu\sim\left\{\frac{\Gamma (5/4)}{\Gamma (1/4)} - 2
\left[ \frac{\Gamma (3/4)}{\Gamma (1/4)}\right]^2\right\}(n+2)^2 4\lambda\, T~~.
\label{flutt-asint}
\end{equation}
In Fig.\ref{fig8} we report the temperature dependence of the time 
average of the Ricci curvature fluctuations, 
$\sigma_\Omega (T)\equiv\langle\delta^2 k_R\rangle_t$. In Fig.\ref{fig9}
we also give a comparison of $\sigma_\Omega$ with the prediction of 
Eq.(\ref{flutt-asint}) for the $O(1)$ model.

The common feature of the three models is that a cusp-like (singular) 
behaviour of the curvature fluctuations is observed in correspondence with 
the phase transition.

Moreover, curvature fluctuations display very smooth energy density dependence, 
or temperature dependence as well, in those systems where no finite order 
phase transition is present (see ref.\cite{CCP}). 
In Fig.\ref{fig10}  we report also $\sigma_\Omega (T)$ in the case 
of 2-d $\varphi^4$ model with $O(2)$ symmetry; in this case a second order 
phase transition is forbidden and actually the system undergoes a 
Kosterlitz-Thouless phase transition. The cusp-like behaviour of curvature 
fluctuations has now disappeared and $\sigma_\Omega (T)$ is a monotonically
increasing function of $T$; visibly, something still happens at the transition 
point ($T_c\simeq 1.5$) so that this case appears to be ``intermediate'' 
between no phase transition at all and a second order phase transition.
Similar results have been found for planar 2-d and 3-d classical 
Heisenberg models \cite{cccp,cecilia_th}  
and in a preliminary investigation of the dual 
(gauge) version of the Ising model in 3-d 
\cite{cecilia_ising}: the cusp-like behaviour of the
curvature fluctuations always shows up when a second order phase transition
is present, and the singular point is located at the critical temperature,
within the numerical accuracy.

In the light of the Riemannian description of Hamiltonian chaos given above, 
we understand why the temperature dependence of the
largest Lyapounov exponent $\lambda_1$ is so peculiar near and at the critical 
temperature (see Figs.\ref{fig4} and \ref{fig5}): 
$\lambda_1(T)$ reflects the ``cuspy'' 
pattern of $\sigma_\Omega (T)$ near $T_c$. In the next Subsection we make
a conjecture about the deep meaning of these singular behaviours shown by
$\lambda_1(T)$ and $\sigma_\Omega (T)$.

As the invariant measure for an autonomous Hamiltonian flow is the 
microcanonical measure on the constant energy surfaces of phase space, our
numerical computations of $\langle k_R\rangle_t$ and  of 
$\langle\delta^2 k_R\rangle_t$ are good estimates of the quantities 
$\Omega_0(T)$ and $\sigma_\Omega (T)$, i.e. microcanonical averages, 
that enter Eqs. (\ref{tau1}) and (\ref{lambda_gauss}). 
The analytic computation of $\lambda_1(T)$ by means of these formulae
yields an unsatisfactory result that overestimates $\lambda_1(T)$ at low
temperatures (though the temperature dependence is correct) and that
steeply increases at high temperatures instead of saturating (before
decreasing again at extremely high $T$). The high temperature result 
appears particularly bad, however this is only due to the asymptotic
growth with $T$ of both $\langle k_R\rangle$ and $\langle\delta^2 k_R\rangle$
--- given by Eqs. (\ref{ricciTinf}) and (\ref{flutt-asint}) and
confirmed numerically --- which has no special meaning for dynamical 
instability. 
The estimate of the decorrelation time-scale of curvature fluctuations along
the geodesics is still somehow rudimental in the above outlined Riemannian
framework, therefore one expects that some improvement is needed 
on this point. As a matter of fact, it is possible to substantially improve 
the theoretical predictions by simply multiplying the decorrelation time scale
$\tau$ of Eq.(\ref{taufinale}) by a constant factor which is model
dependent and different below and above $T_c$. Moreover, at high temperatures,
in computing $\tau_1$ and $\tau_2$ given in Eq.(\ref{tau1}),
we have subtracted to
$\Omega_0(T)$ and $\sigma_\Omega (T)$ their respective asymptotic behaviours
given by Eqs. (\ref{ricciTinf}) and (\ref{flutt-asint}). The analytic
predictions for $\lambda_1(T)$ are now in very good agreement with numeric
results with the exception of the critical region, where something is 
apparently still lacking. The results are reported in Figs.\ref{fig11},
\ref{fig12} and \ref{fig13} where it is well evident that the best agreement 
between theory and numerical experiments is obtained in the $O(1)$ case; a
very good agreement is still present at low temperatures for the $O(2)$ model
and it becomes poorer in the $O(4)$ model. The comparison at $T>T_c$ suffers --
in the cases of $O(2)$ and $O(4)$ -- of a restricted range of temperature
values (we focused our attention only to the transition region because of
the already mentioned problems) where subtracting to $\Omega_0(T)$ and 
$\sigma_\Omega(T)$ their asymptotic values is less meaningful.

However it is not out of place to remind that the theoretical computation of
Lyapounov exponents is not a routine task at all, and that the approach reported
here is at present the only theoretical method available to cope with the
computation of $\lambda_1$. 
What is important here is that with some simple and reasonable 
adjustment the above sketched analysis still applies
and yields good results. Refinements of the geometrical theory of chaos are
beyond the aim of the present work, rather we are interested in using it as it
is at present to get a hold of the deep origin of the peculiarities of 
the dynamics at a phase transition.

\subsection{A topological conjecture}

We shall now try to grasp the possible significance of the above reported 
cusp-like, thus possibly singular, behaviour of the curvature fluctuations
at the transition point for the $\varphi^4$ lattice systems.
As a first step toward this goal, we shall try to reproduce such a peculiar
behaviour of curvature fluctuations in abstract geometric models.
A preliminary step in this direction was already presented in Ref.
\cite{cccp}, applied to the case of planar spin models.

The choice of a geometric {\it toy model} stems from the following
considerations. Weakly and strongly chaotic geodesic flows can ``live''
on homologically trivial manifolds, i.e., on manifolds that are diffeomorphic
to an $N$-sphere, in other words, non-trivial topology is not necessary to
make chaos, conversely, a sudden topological 
change in a family of manifolds can abruptly 
affect their geometric properties and the degree of chaoticity
of geodesic flows. Therefore, let us consider 
--- for instance --- the two families of surfaces of revolution immersed
in ${\Bbb R}^3$ defined as follows:
\begin{mathletters}
\begin{equation}
{\cal F}_\varepsilon   =  
(f_\varepsilon (u) \cos v, f_\varepsilon (u) \sin v, u) ~,
\label{F}
\end{equation}
\begin{equation}
{\cal G}_\varepsilon  = 
(u \cos v, u \sin v, f_\varepsilon (u))~,
\label{G}
\end{equation}
\end{mathletters}
where
\begin{equation}
f_\varepsilon (u) = \pm \sqrt{\varepsilon + u^2 - u^4}~,\qquad
\varepsilon \in [\varepsilon_{\text{min}},+\infty)~,
\end{equation}
and $\varepsilon_{\text{min}} = -\frac{1}{4}$. Some members of the two families 
are depicted in Fig. \ref{figfamilies}. In both cases there exists a
critical value of the parameter $\varepsilon$, $\varepsilon_c = 0$,
corresponding to a change in the {\em topology} of the surfaces. In particular
the manifolds ${\cal F}_\varepsilon$ 
are diffeomorphic to a torus ${\Bbb T}^2$ for
$\varepsilon < 0$ and to a sphere ${\Bbb S}^2$ for
$\varepsilon > 0$. In the other case, one has instead a change in the number
of connected components: the manifolds 
${\cal G}_\varepsilon$ are diffeomorphic to {\em two} spheres
for $\varepsilon < 0$ 
and to one sphere for $\varepsilon > 0$. Computing the Euler-Poincar\'e 
characteristic $\chi$ one finds 
$\chi({\cal F}_\varepsilon)  =  0$ if $\varepsilon < 0$, and 
$\chi({\cal F}_\varepsilon)  =  2$ otherwise, while 
$\chi({\cal G}_\varepsilon)$ equals either $4$ or $2$ when $\varepsilon$ is
respectively negative or positive. 
Let us now compute the $\varepsilon$-dependence  of the average 
curvature properties of these surfaces as 
$\varepsilon \to \varepsilon_c$. Let $M$ belong to one of the two families
under investigation.
The gaussian curvature $K$ is given by \cite{Spivak}
\begin{equation}
K= {{x^\prime (x^{\prime\prime}y^\prime - x^\prime y^{\prime\prime})}\over
{y (x^{\prime 2} + y^{\prime 2})^2}}
\label{curvgauss}
\end{equation}
where the functions $x(u)$ and $y(u)$ represent the coefficients of the general
form $M(u,v)=(y(u)\cos v, y(u) \sin v, x(u))$ of parametrized surfaces of 
revolution, and the prime denotes differentiation with respect to $u$.
Now the fluctuations of $K$ are computed as follows 
\begin{equation}
\sigma^2 = \langle K^2 \rangle - \langle K \rangle^2 = 
A^{-1}\int_M K^2 \, dS - \left(A^{-1} \int_M K \, dS\right)^2~,
\label{rms_modelli}
\end{equation}
where $A$ is the area of $M$ and $dS$ is the invariant surface element.
Both families of surfaces, in spite of having very different curvature
properties on the average\footnote{For instance, 
$\langle K \rangle(\varepsilon) = 0$ in the 
${\cal F}_\varepsilon$ case as $\varepsilon < 0$, 
while the same average curvature is positive and diverging
as $\varepsilon \to 0$ for ${\cal G}_\varepsilon$.}, 
exhibit a singular behaviour in the curvature fluctuation $\sigma$ as
$\varepsilon \to \varepsilon_c$, as shown in Fig. \ref{figmodels}.

These results suggest --- at a heuristic level --- that, 
from the point of view of the
geometric description of the dynamics, a phase transition might correspond
to a {\em topology change} in the manifold underlying the motion.
The relevance of topological concepts for the theory of phase transitions
has been already emphasized (see Ref. \cite{Rasetti}) though in a more
abstract context. Here we suggest that topological aspects of phase
transitions might also concern the manifolds that are ``just behind'' dynamics,
and not only those deep mathematical objects that are involved in Ref. 
\cite{Rasetti}. In our opinion this subject 
deserves further investigation to go beyond the heuristic level. In fact
the study of dynamics and of its geometric and topologic counterparts  
could eventually lead to a better understanding
of the nature of ergodicity breaking and thus of the very nature
of phase transitions.

\section{Concluding remarks}
\label{sec_comments}
Let us now summarize the main points of the present work and comment about their
meaning.

By studying some classical lattice $\varphi^4$ models that undergo second order
or Kosterlitz-Thouless phase transitions, it has been found that the natural
microscopic dynamics --- derived from the Hamiltonian functions of these
systems ---  clearly reveals the presence of the phase transition. The
invariant measure of Hamiltonian dynamics is the microcanonical measure,
equivalent --- in the thermodynamic limit ---  to the canonical measure which is
sampled by usual MonteCarlo algorithms. Therefore one could argue that it is
not surprising that Hamiltonian dynamics yields the same results of a 
MonteCarlo stochastic dynamics.
As a matter of fact using Hamiltonian dynamics just to
sample the microcanonical measure would not be so interesting, whereas 
the important point raised by the present work is that Hamiltonian dynamics
brings about new observables and a new framework to tackle phase transitions. 
Mainly Lyapounov exponents are the new observables intrinsic to the dynamics,
and the differential-geometric treatment of dynamical instability is the new
framework. 

As well as thermodynamic observables, dynamic and geometric 
observables are sensitive to a second order phase transition which can be
recognized through their peculiar ``non-smooth'' behaviours.
The common wisdom on phase transitions suggests that non-smooth behaviours
of any observable are expected near the transition point, as a consequence of
the tendency of the measure to become singular at $T=T_c$ in the limit
$N\rightarrow\infty$. In the light of our results we suggest that a deeper 
explanation might be possible: a major topological change of the ``mechanical''
manifolds could be the common root of the peculiar behaviours of both dynamic 
and thermodynamic observables in presence of a phase transition. Here topology
is meant in the sense of the De Rham's cohomology.

On a purely phenomenological ground it might be surprising that the largest
Lyapounov exponent, which measures an average {\it local} property of the
dynamics, is sensitive to a {\it collective}, and therefore global, phenomenon
like a phase transition. 
In fluids, for example, it is well evident that molecular chaos
has nothing to do with the macroscopic patterns of the velocity field. It is
even possible to have chaotic motions of fluid droplets (Lagrangian chaos)
in presence of regular Eulerian velocity fields (i.e. in laboratory reference 
frame).

However, within the Riemannian framework outlined in the preceding Sections,
Lyapounov exponents appear tightly related to the geometry of the ``mechanical''
manifolds, and geometry dramatically changes in presence of a major change of
topology. Thus our topological conjecture seems to naturally account for this
-- at first sight counterintuitive -- sensitivity of the largest Lyapounov
exponent to a macroscopic collective phenomenon.

It is worth mentioning here that, at the best of our knowledge, there is only 
another framework where Lyapounov exponents can be, at least in principle, 
analytically computed. This is a field-theoretic framework, already mentioned
in the Introduction \cite{reuter,gozzi1,gozzi2}, based on a path-integral 
formulation of classical mechanics, where Lyapounov exponents are seen as 
expectation values of suitable operators.
There are many interesting points in this framework that could probably reveal
a fertile relationship with the Riemannian geometric approach which is behind
our present work. Let us mention some of them: ergodicity breaking, which - as 
we discussed in the Introduction - is a more general concept than symmetry
breaking, in the field-theoretic context appears related to a supersymmetry 
breaking, moreover this supersymmetry breaking can occur also at finite $N$
\cite{reuter}; Lyapounov exponents turn out related to mathematical objects that 
have many analogies with definitions and concepts of the de Rham's cohomology 
theory \cite{gozzi2} what might be useful in future investigations about the 
relation between Lyapounov exponents and topology at  a phase transition.

In conclusion we believe that the dynamic approach, 
besides the conceptual aspects mentioned above,
could  contribute to complement the standard approaches of statistical
mechanics to the description of phase transitions, and it might hopefully be
particularly helpful in those cases where these standard methods may 
encounter some difficulty, as is the case of disordered and frustrated systems,
polymers in the continuum, 
lattice gauge theories where no symmetry-breaking transition
occurs.

\acknowledgments

We thank S. Caracciolo, E. G. D. Cohen, R. Livi, and M. Rasetti for 
many fruitful discussions. This work has been carried out under the
EC program Human Capital and Mobility, contract N$^\circ$ UE 
ERBCHRXCT940579 and N$^\circ$ OFES 950200.

\begin{figure}
\caption{Binder cumulants $g(N,\varepsilon )$ {\it vs} energy density 
$\varepsilon$  at different values $N$ of the lattice sites. {\it a)} the 
$O(1)$ case; {\it b)} the $O(2)$ case; {\it c)} the $O(4)$ case. Open circles
refer to $N=4\times 4\times 4$, full triangles refer to $N=6\times 6\times 6$,
and full circles refer to $N=8\times 8\times 8$.  }
\label{fig1}
\end{figure}

\begin{figure}
\caption{Temperature $T$ (twice the average kinetic energy per particle) is 
plotted {\it vs} energy density $\varepsilon$. Results of the $O(1)$, $O(2)$
and $O(4)$ models are represented by full circles, open circles and open
triangles respectively. Temperatures and energy densities of each model
are scaled by the corresponding critical values obtained by means of Binder 
cumulants. The dashed line is a guide to the eye. 
Lattice size: $N=8\times 8\times 8$}
\label{fig2}
\end{figure}

\begin{figure}
\caption{Specific heat per degree of freedom {\it vs} scaled temperature
$T/T_c$.~$c_V=C_V/Nn$ and $C_V$ is computed according to 
Eq.(\protect\ref{formulalpv}). Symbols: full circles for $O(1)$; open circles
for $O(2)$ and open triangles for $O(4)$. Lattice size: $N=8\times 8\times 8$.}
\label{fig3}
\end{figure}

\begin{figure}
\caption{The largest Lyapounov exponent $\lambda_1$ is plotted {\it vs}
temperature for the $O(1)$ model. A ``non-smooth'' feature at $T=T_c$ is
well evident. Lattice size: $N=8\times 8\times 8$. }
\label{fig4}
\end{figure}

\begin{figure}
\caption{Synopsis of $\lambda_1(T)$ obtained for the $O(1)$ model (full 
circles), for the  $O(2)$ model (open circles) and for the  $O(4)$ model
(open triangles). Lattice size: $N=8\times 8\times 8$. }
\label{fig5}
\end{figure}

\begin{figure}
\caption{Reduced average Ricci curvature $\kappa=(\langle k_R\rangle -2Jd)/
(\langle k_R(T=0)\rangle -2Jd)$ {\it vs} $T/T_c$. Ricci curvature is so
reduced in order to facilitate the comparison between the different models.
Symbols: full circles for $O(1)$; open circles
for $O(2)$ and open triangles for $O(4)$. Lattice size: $N=8\times 8\times 8$.}
\label{fig6}
\end{figure}

\begin{figure}
\caption{ $O(1)$ model. The average Ricci curvature $\langle k_R\rangle$ is 
plotted {\it vs} $T$ for a wide temperature range. The dashed horizontal line
represents the integrable limit behaviour of $\langle k_R\rangle (T)$
predicted by Eq.(\protect\ref{ricciT=0}) and actually attained at low 
temperature by the average Ricci curvature computed for the $O(1)$ model
according to Eq.(\protect\ref{riccifi4}). Solid line represents the high 
temperature asymptotic behaviour of $\langle k_R\rangle(T)$ predicted by
Eq.(\protect\ref{ricciTinf}), again pertaining an integrable limiting behaviour
of the model.  }
\label{fig7}
\end{figure}

\begin{figure}
\caption{Average Ricci curvature fluctuations $\sigma_\Omega$ {\it vs} $T/T_c$.
The ``cuspy'' behaviour is well evident at $T\simeq T_c$. From top to bottom:
$O(4)$, $O(2)$ and $O(1)$ results. The cusp appear to soften at increasing
dimension $n$ of the symmetry group $O(n)$.  }
\label{fig8}
\end{figure}

\begin{figure}
\caption{Average Ricci curvature fluctuations $\sigma_\Omega$ {\it vs} $T$
for the $O(1)$ model reported for a wide range of temperature. Solid line
represents the high temperature asymptotic value given 
by Eq.(\protect\ref{flutt-asint}).  }
\label{fig9}
\end{figure}

\begin{figure}
\caption{Average Ricci curvature fluctuations $\sigma_\Omega$ {\it vs}
temperature for the $O(2)$ model on a square lattice ($d=2$) of 
$N=30\times 30$ sites. The cusp is now absent and $\sigma_\Omega (T)$ is a
monotonously increasing function. Around $T\simeq 1.5$, on the basis of the
temperature behaviour of other observables, the system is supposed to undergo 
a Kosterlitz-Thouless phase transition and, correspondingly, we can observe 
a change in the shape of $\sigma_\Omega (T)$. Here $J=1$, $\lambda =4$ and 
$m^2=10$. }
\label{fig10}
\end{figure}

\begin{figure}
\caption{The numerical largest Lyapounov exponent $\lambda_1$ (open circles)
is plotted {\it vs} $T$ for the $O(1)$ model and is compared to the analytic
prediction of Eq.(\protect\ref{lambda_gauss}) (full circles). The vertical 
solid line marks the transition temperature. The correlation time scale $\tau$
is given by Eq.(\protect\ref{taufinale}); $\tau$ is rescaled by a constant
factor equal to $0.65$ at $T <T_c$, and by a factor $1.1$ at $T>T_c$.  }
\label{fig11}
\end{figure}

\begin{figure}
\caption{The numerical largest Lyapounov exponent $\lambda_1$ (open circles)
is plotted {\it vs} $T$ for the $O(2)$ model and is compared to the analytic
prediction of Eq.(\protect\ref{lambda_gauss}) (full circles). The vertical 
dashed line marks the transition temperature. Here $\tau$ is rescaled 
by a constant factor equal to $3$ at $T <T_c$, and by a factor $0.7$ 
at $T>T_c$.  }
\label{fig12}
\end{figure}

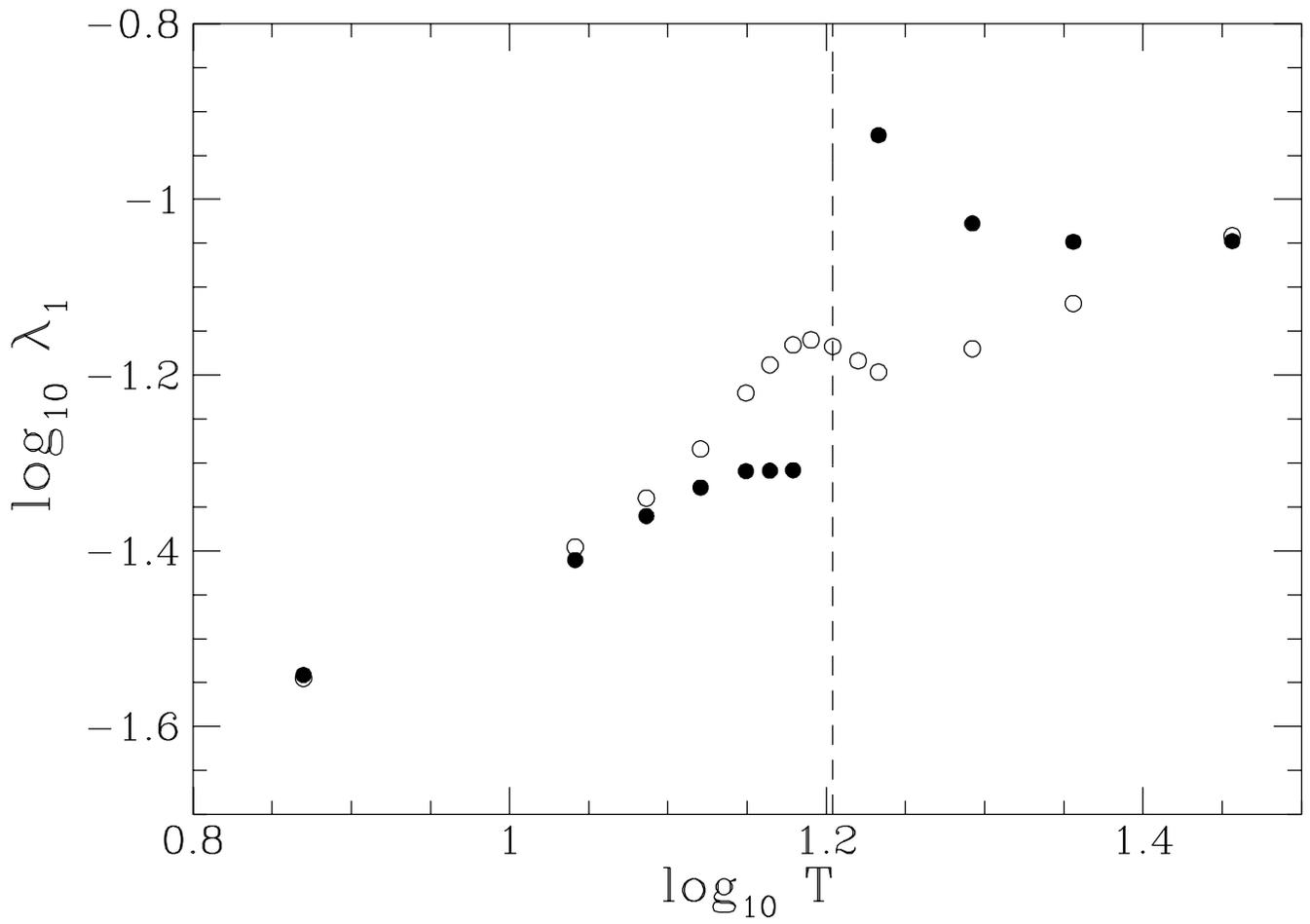
\begin{figure}
\caption{The numerical largest Lyapounov exponent $\lambda_1$ (open circles)
is plotted {\it vs} $T$ for the $O(4)$ model and is compared to the analytic
prediction of Eq.(\protect\ref{lambda_gauss}) (full circles). The vertical 
solid line marks the transition temperature. Here $\tau$ is rescaled by 
a constant factor equal to $5.5$ at $T <T_c$, and by a factor $0.6$ at 
$T>T_c$. }
\label{fig13}
\end{figure}

\begin{figure}
\caption{Some representatives of the two families of surfaces 
${\cal F}_\epsilon$ and ${\cal G}_\epsilon$ defined in Eqs.(\protect\ref{F})
and (\protect\ref{G}) respectively. Each family is divided into two
subfamilies by the critical surface corresponding to $\epsilon_c=0$ (middle 
members in the picture). Members of the same subfamily are diffeomorphic,
whereas the two subfamilies are not diffeomorphic between them.  }
\label{figfamilies}
\end{figure}

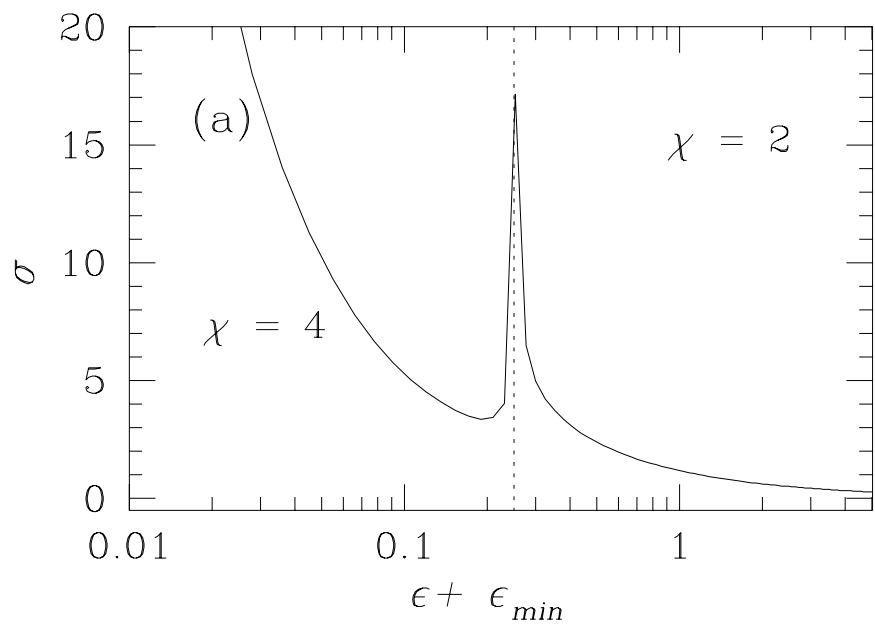
\begin{figure}
\caption{The second moment of the gaussian curvature of the surfaces
${\cal F}_\epsilon$ and ${\cal G}_\epsilon$ is plotted {\it vs} $\epsilon$.
$\sigma$ is defined in Eq.(\protect\ref{rms_modelli}), $\epsilon$ is shifted
by $\epsilon_{min}=0.25$ (see text) for graphical reasons. {\it (a)} refers to
${\cal G}_\epsilon$ and {\it (b)} refers to ${\cal F}_\epsilon$. The cusps
appear at $\epsilon =0$ where the topological transition takes place. }
\label{figmodels}
\end{figure}

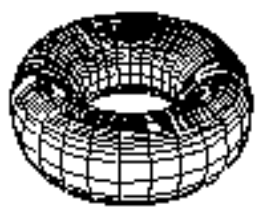
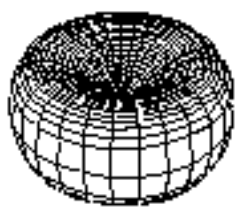
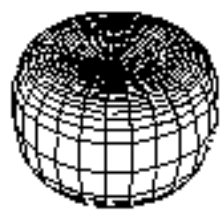
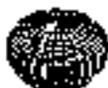
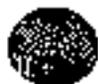
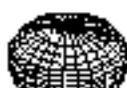
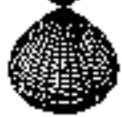
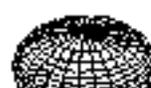
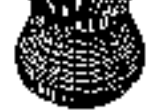
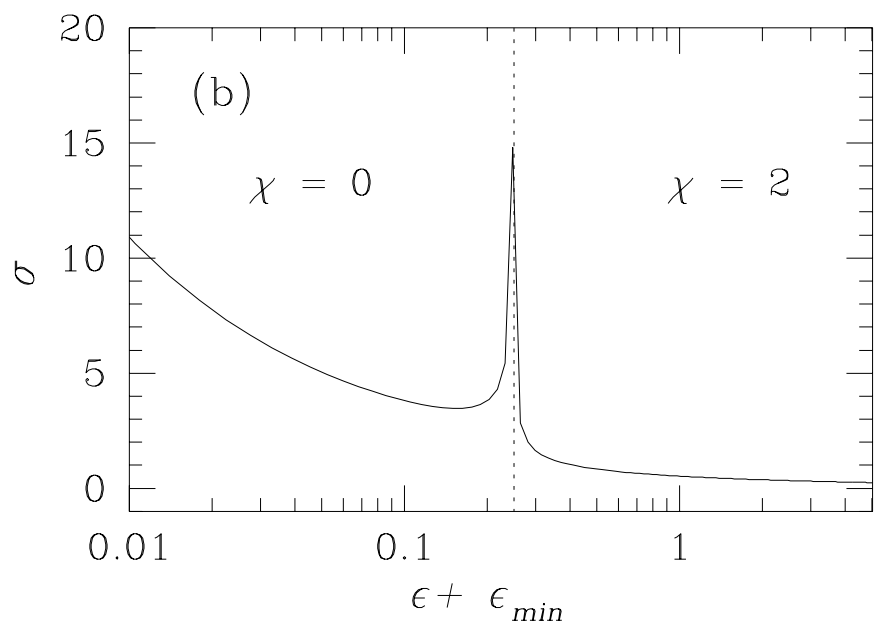
\end{document}